\documentstyle[epsf]{mn}

\newcommand{\etal}{{et~al.}}
\newcommand{\ei}{{\it Einstein}}
\newcommand{\ro}{{\sl ROSAT}}

\newcommand{\eu}{{\sl EUVE}}
\newcommand{\iu}{{\sl IUE}}
\newcommand{\hs}{{\sl HST}}
\newcommand{\HST}{{\sl HST}}

\newcommand{\hip}{{\sl Hipparcos}}
\newcommand{\fuse}{{\sl FUSE}}

\newcommand{\secpoint}{''\mskip-7.6mu.\,}
\newcommand{\degpoint}{^\circ\mskip-7.0mu.\,}
\newcommand{\iue}{{\it IUE}}

\title[Resolving Sirius-like binaries]
{Resolving Sirius-like binaries with the {\it Hubble Space
Telescope}\thanks{Based on observations with the NASA/ESA {\it Hubble Space
Telescope}, obtained at the Space Telescope Science Institute, which is
operated by AURA, under NASA contract NAS 5-26555.}} 
\author[M.A. Barstow et al.]{
M.A. Barstow$^{1}$, Howard E. Bond$^{2}$, M.R. Burleigh$^1$, 
and J.B. Holberg$^3$
\\
$^1$ {\it Department of Physics and Astronomy, University of Leicester,
University Road, Leicester LE1 7RH, UK}\\
$^2$ {\it Space Telescope Science Institute, 3700 San Martin Dr.,
Baltimore, MD 21218, USA}\\
$^3$ {\it Lunar and Planetary Laboratory, University of Arizona, Tucson, 
AZ 85721, USA}
}

\begin{document}

\label{firstpage}

\maketitle

\begin{abstract}

We present initial results from a {\it Hubble Space Telescope\/}
ultraviolet imaging 
survey of stars known to have hot
white-dwarf companions which are unresolved from the ground.  The hot
companions,
discovered through their EUV or UV emission, are
hidden by the overwhelming brightnesses of the
primary stars at visible wavelengths. Out of 17 targets observed, we have 
resolved eight of them with the Wide Field Planetary Camera~2, using various 
ultraviolet filters.
Most of the implied orbital periods for the resolved systems
are hundreds to thousands of years,
but in at least three cases (56 Persei, $\zeta$~Cygni, 
and RE~J1925$-$566) it should
be possible to detect the
orbital motions within the next few years, and they may eventually yield new 
dynamically determined masses for the white-dwarf components.
The 56~Persei and 14~Aurigae systems are found to be quadruple and quintuple, 
respectively, including the known optical components as well as the newly 
resolved white-dwarf companions. The mild barium star $\zeta $~Cygni,
known to have an 18 year spectroscopic period, is marginally resolved.
All of these new resolved
Sirius-type binaries
will be useful in determining gravitational redshifts and masses 
of the white-dwarf components.

\end{abstract}

\begin{keywords} stars: binaries 
-- stars: white dwarfs -- ultraviolet: stars.
\end{keywords}
 
\section{Introduction}

White dwarfs (WDs) 
represent the evolutionary end point for most stars,
since it is believed that all stars with initial
masses below $\approx$$8\,M_\odot$ become WDs.
An understanding of the evolution of WDs is important in several 
contexts.
For example, the loss of the stellar envelope in
the form of a planetary nebula, during the transition from red giant to hot
WD, recycles stellar material back into the interstellar
medium. Imprinted on the WD luminosity function is the history of
star formation in our Galaxy, and the cut-off at $\approx$3500~K, below which
no WDs are known, places a limit of 8--10 Gyr on the age of
the disk (e.g. Oswalt \etal\ 1996; Leggett, Ruiz, \& Bergeron 1998). 

Theoretical studies have
led to the formulation of the mass-radius relation and
Chandrasekhar mass limit for zero-temperature WDs
(e.g. Chandrasekhar 1931, 1935; Hamada \& Salpeter 1961), and later to
inclusion of the effects of the actual WD temperature due to the
presence of a non-degenerate envelope (Wood 1992) and of variations in the core
composition (Wood 1995).  Direct tests of the theory are still rare, 
however,
due to 
the difficulty of obtaining
accurate measurements of the radii and masses of known WDs. $T_{\rm
eff}$ and $\log g$ can be measured from the Balmer-line profiles in optical
spectra, using theoretical model atmospheres 
(Holberg \etal\ 1985; Kidder 1991). Bergeron, Saffer, \&
Liebert (1992) used this technique to determine
the
mass distribution of the population of single H-rich DA WDs, by
combining the observational results with the theoretical mass-radius relation
of Wood (1992). 

A combination of Balmer-line measurements and parallax data can provide an
independent means of obtaining WD masses and radii and, therefore, a
new route for examining the evolutionary models. Until recently, 
most parallax measurements of WDs have not been of
sufficient accuracy to provide meaningful constraints, but the improvement in
ground-based observations (Monet \etal \ 1992) and the
availability of space-based parallaxes from \hip\ (ESA 1997) has
improved the situation. Recent results suggest that the
masses and radii are generally consistent with expectations (Vauclair \etal\
1997; Barstow \etal\ 1997; Provencal \etal\ 1998;
Holberg \etal \ 1998). However, most of this
work is restricted to the narrow range of masses near $0.6\,M_\odot$,
encompassed in the population of isolated WDs. Furthermore, even
with the new parallaxes, the overall uncertainties remain 
too large to provide a
really thorough examination of the differences between theoretical models
which make differing assumptions about the core and envelope
compositions. 

Observations of WDs in binaries potentially represent a more direct
test of the evolutionary models, since a dynamical WD mass can
be determined 
purely from the orbital elements
of the system. In practice, however, very
few such systems have been available to
be studied in sufficient detail to make such comparisons; only Sirius, 
Procyon (Girard \etal \ 2000), and a handful of other binary 
systems have provided useful directly measured WD masses to 
date. 

This situation is likely to 
improve dramatically in the next few years, because
the \ro\ Wide Field Camera
(WFC) and \eu\ sky surveys have revealed
a number of new sources in which strong
EUV emission is detected from a previously unknown hot, but optically
faint, companion of a
visually bright star. Several similar systems have also been discovered
serendipitously in the TD-1 survey data (Landsman \etal\ 1996) and in \iu\
observations of individual stars (e.g. B\"ohm-Vitense 1980, 1993). 

This interesting, but rather disparate, group of objects has been the
subject of a series of studies (e.g. Barstow \etal\ 1994; Vennes \etal\
1995; Genova \etal\ 1995; Burleigh, Barstow, \& Fleming 1997; Burleigh,
Barstow, \& Holberg 1998; Vennes, Christian, \& Thorstensen 1998).  
Some of them should be new
members of the class of wide binaries represented by 
Sirius,
comprising a normal dwarf or subgiant and a WD
companion, and they
present an opportunity to extend the sample of
well-studied binary systems. 
Others may belong to much closer binaries, which 
have passed through a common-envelope interaction; however,
with the exception of HR~8210 (a spectroscopic binary with an orbital period 
of 21.7~days), radial-velocity
measurements have not revealed any of the
objects to have a sufficiently short binary
period (e.g. Vennes \etal\ 1998). 

None
of these systems have been resolved using ground-based imaging, down to 
typical resolution limits of
$0\secpoint5$ to $1\secpoint0$,
but a major complication is the overwhelming
brightness of the primary in the visible band. Since
the difference in visual
brightness between primary and WD can be 5~mag or more, a
WD could easily be hidden in the wings of the image
of the primary. This makes the effective resolution several times worse
than what can be achieved with smaller magnitude differences.

The {\it Hubble Space Telescope\/} (\HST\/) and its Wide Field Planetary 
Camera~2 (WFPC2)
provide the answer to these problems.
WFPC2 allows imaging in the ultraviolet, where the
brightnesses of both components become comparable. Moreover, WFPC2
delivers diffraction-limited images at UV wavelengths of
$\approx0\secpoint08$ FWHM,
making it possible to resolve binary components with separations down to
about this amount.

In this paper we report the initial results of an \HST\/
``snapshot'' survey for 
resolved Sirius-like
binary systems.  Out of 17 targets observed to date, we have resolved 8 of 
them.  In the following sections we discuss our analysis of the WFPC2
images, the nominal orbital periods for the resolved systems, and the limits
on the orbital periods for those binaries that we were unable to resolve.
In addition, we discuss the results for each individual object in the
context of their known properties from earlier observations.

\section{Target properties}

Table~\ref{binpar} lists the physical parameters of the 17
systems that we have 
observed to date with WFPC2. 
The data are drawn from the references cited above, as well 
as the SIMBAD database, and 
include the spectral types and $V$ magnitudes of the optical primary stars, the 
distances to the systems 
(mostly derived from the \hip\/ parallaxes, or in six cases estimated 
from the spectral types and magnitudes), and the effective temperatures and 
approximate $V$ magnitudes of the WD components.
The parameters
for the WDs
have been derived mainly from their Lyman~$\alpha$ profiles by
ourselves and other authors 
(see references in Table~\ref{binpar}).
However,
such measurements suffer from known ambiguities. Consequently, where
possible, these parameters have been further constrained by soft X-ray/EUV
photometry and spectroscopy, together with knowledge of the primary distance.
The primary stars have
spectral types ranging from A1 to K2, and with a few exceptions 
they are dwarfs 
or subgiants. 

\begin{table*}
\caption{Physical parameters for the target binaries. Stars marked
($^*$) have a \hip\/ parallax measurement; other distances are estimated from 
the spectral type and magnitude} 
\label{binpar}
\begin{tabular}{lllllrll}
Star & Alt. Name & Spectral   & $V_{\rm pri}$ & $d$  & $T_{\rm WD}$ &
$V_{\rm WD}$ & Ref.\\
     & & Type       &               & (pc) & (K)          & (est.) & \\

HD 2133$^*$ & RE~J0024$-$071 & 
F7 V & 9.6 & 140 & 27,000 & 15.6 & Burleigh \etal\ 1997\\
BD +08$^\circ$102 & RE~J0044$+$093 & 
K2 V & 10.2 & 62 & 26,000 & 14.2 & Vennes \etal\ 1998\\
HD 15638$^*$ & CD~$-$61$^\circ$431 & 
F6 V & 8.8 & 199 & 45,000 & 14.7 & Vennes \etal\ 1998 \\
HD 18131$^*$ & BD~$-$05$^\circ$541 & 
K0 IV-III & 7.2 & 104 & 34,000 & 14.5 & Vennes \etal\ 1998 \\
MS 0354.6$-$3650 & EUVE~J0356$-$366 & 
G2 V & 12.4 & 400 & 52,000 & 17.4 & Christian \etal\ 1996\\
HR 1358$^*$ & HD 27483 & 
F6 V & 6.2 & 46 & 22,000 & 14.5 & Burleigh \etal\ 1998 \\
56 Per$^*$ & HD 27786 & 
F4 V & 5.8 & 42 & 14,500 & 15.0 & Landsman \etal\ 1996 \\
63 Eri$^*$ & HR 1608 & 
K0 IV & 5.4 & 55 & 27,000 & 13.1 & Vennes \etal\ 1998\\
14 Aur C$^*$ & HD 33959C & 
F4 V & 8.0 & 82 & 42,000 & 14.1 & Holberg \etal\ 1999 \\
HR 3643$^*$ & HD 78791 & 
F7 II & 4.5 & 139 & 29,000 & 14.5 & Landsman \etal\ 1996\\
HD 90052 & BD~$+$27$^\circ$1888 & 
F0 V & 9.6 & 250 & 33,000 & 15.0 & Burleigh \etal\ 1997\\
$\beta$ Crt$^*$ & HR 4343 & 
A1 III & 4.5 & 82 & 31,000 & 13.9 & Vennes \etal\ 1998\\
RE J1309+081 & EUVE~J1309$+$082 & 
F9 V & 11.5 & 275 & $>$22,000 & 16.0 & this paper \\
RE J1925$-$566 & EUVE~J1925$-$565 & 
G7 V & 10.6 & 110 & 50,000 & 14.7 & Vennes \etal\ 1998 \\
$\zeta$ Cyg$^*$ & HR 8115 & 
G8 IIIp & 3.2 & 46 & 12,000 & 13.2 & Dominy \& Lambert 1983\\
HR 8210$^*$ & IK Peg & 
A8 Vm & 6.1 & 46 & 35,000 & 14.4 & Vennes \etal\ 1998 \\
HD 223816 & CD~$-$71$^\circ$1808 & 
F8 V & 8.8 & 92 & 69,000 & 14.2 & Vennes \etal\ 1998 \\

\end{tabular}
\end{table*}

\section{WFPC2 Observations}

All observations of the binary systems were conducted between 1999 July 17
and 2000 June 10, as part of our Cycle~8 ``snapshot'' 
programme (GO-8181 - the programme remains active at the time of writing, 
but it is now unlikely 
that any
additional targets may be observed). In each case we
placed the target on the Planetary Camera CCD chip of the WFPC2
camera on board \HST\null. 
A single UV filter was chosen, for each system, to give
approximately equal fluxes from the two components, based on knowledge of the
primary spectral type and temperature of the WD\null. This optimised the
ability to separate the stars in the image, if resolvable. Each complete
observation consisted of a pair of equal
exposures (to facilitate removal of cosmic-ray hits from the images). The
exposure times were calculated where possible
to achieve the maximum possible signal-to-noise, just short of saturation of
the CCD by the brighter component in that particular band, but were limited 
to a maximum of 500~s. In most cases the
exposures did prove to be near optimal, but a few objects had a few saturated
pixels at the centre of the brighter component's image.  In two cases 
(HR~1358 and 56~Per) there 
was a guide-star acquisition failure, but the target was nevertheless located 
within the PC field of view and the observation was successful although 
obtained only under gyro control.

Table~\ref{obs}
summarises the observing programme, with the final column indicating whether 
or not the binary was resolved. The filter names indicate their nominal 
effective wavelengths in nm.

\begin{table*}
\caption{Details of WFPC2 observations}
\label{obs}
\begin{tabular}{llccl}
Star     &  Obs.\ Date & Filter & Exposure (s) & Comment \\

HD 2133 & 1999 Aug 23 & F255W & $2\times 180$ & resolved \\
BD +08$^\circ$102 & 1999 Oct 16 & F300W & $2\times18$ & unresolved\\
HD 15638  & 1999 Oct 5 & F218W & $2\times 70$ & unresolved\\
HD 18131 & 1999 Aug 18 & F255W & $2\times80$ & unresolved\\
MS 0354.6$-$3650 & 2000 May 25 & F255W & $2\times500$ & resolved\\
HR 1358 & 1999 Jul 17 & F185W & $2\times80$ & resolved\\
56 Per  & 1999 Oct 17 & F170W & $2\times120$ & 4 components\\
63 Eri  & 1999 Aug 16 & F255W & $2\times35$ & unresolved\\
14 Aur C  & 1999 Sep 23 & F185W  & $2\times70$ & resolved \\
HR 3643 & 1999 Sep 18 & F185W & $2\times100$ & unresolved\\
HD 90052 & 2000 May 31 & F185W & $2\times120$ & unresolved\\
$\beta$ Crt & 2000 Jun 10 & F160BW & $2\times2.6$ & unresolved\\
RE J1309+081 & 1999 Aug 17 & F218W & $2\times 500$ & unresolved\\
RE J1925$-$566 & 1999 Aug 28 & F300W & $2\times26$ & resolved\\
$\zeta$ Cyg & 2000 May 23 & F218W & $2\times60$ & resolved\\
HR 8210  & 1999 Jul 19 & F170W & $2\times8$ & unresolved\\
HD 223816  & 1999 Sep 28 & F255W & $2\times50$ & resolved\\

\end{tabular}
\end{table*}

\section{Analysis of the WFPC2 images}

All images were processed through the standard WFPC2
pipeline provided by STScI\null.  
We inspected each image carefully, and in eight cases the binary was either 
clearly resolved into two (or more) stars, or (for $\zeta$~Cyg)
had an elongated, partially resolved image.  For the remaining nine systems we 
measured the FWHM, ellipticity, and position angle of the elliptical fit
to the image (using IRAF's\footnote{
IRAF is distributed by the National Optical Astronomy Observatories,
which are operated by the Association of Universities for Research
in Astronomy, Inc., under cooperative agreement with the National
Science Foundation.} 
{\it imexamine\/} routine). This procedure
revealed no 
evidence of resolution of the images, although it did show significant 
variations in the FWHM and ellipticity from object to object, attributed to 
telescope ``breathing'' (i.e., small changes in focus due to the varying 
thermal environment as the spacecraft orbits the Earth).

\subsection{Resolved systems}

For the resolved binaries, we measured the component locations and
relative magnitudes, again using the IRAF
{\it imexamine\/} routine.  In order to convert the positions of the stellar 
components into the true
separation and position angle, we need to correct for the 
wavelength-dependent geometric distortions known to exist
in the PC camera.  

Wavelength-dependent formulae that correct for the geometric distortion have 
been calculated by Trauger et al.\ (1995) based on a detailed 
analysis of the camera optics, and are adopted here. These 
formulae convert measured $(x,y)$ pixel
coordinates at a given wavelength
into geometrically corrected 
coordinates at a wavelength of 5550~\AA\null. In implementing this 
correction, we take the effective wavelength of each filter to be given by 
the filter name, and we have not attempted to correct for the
small effect of
the different 
amounts of flux transmitted by the filter red-leaks for the blue and red 
components of the binaries.  We then convert $(x_{\rm 
corr},y_{\rm corr})$ to 
seconds of arc, using an adopted plate scale of $0\secpoint04557 \pm 
0\secpoint00002 \, \rm pixel^{-1}$ in the F555W filter (Trauger et al.\ 
1995).  

There have been relatively few detailed empirical measurements done to date
to confirm the Trauger et al.\ predicted formulae for the WFPC2
UV filters. We therefore downloaded a set of 
UV images of the star cluster NGC~2100 and the elliptical galaxy NGC~205 from 
the STScI archive, 
measured the separations of several pairs of stars in images taken in
each of the UV filters, 
and corrected the separations for the wavelength-dependent geometry as 
described above.  The results showed very small, but significant,
residual filter-dependent
differences in the measured 
stellar separations.  We assumed these differences to be correctable by small 
scale factors, as given in Table~\ref{platescale} (the sense of the 
corrections
being that the measured separations should be multiplied by these
scale factors to yield the true separations).

\begin{table}
\caption{WFPC2 PC chip scale correction factors}
\label{platescale}
\begin{tabular}{ll}
Filter & Correction Factor \\

F160BW  & $0.9982 \pm 0.0004 $ \\
F170W   & $1.0009 \pm 0.0001 $ \\
F185W   & $1.00036 \pm 0.00004 $ \\
F218W   & $1.0001 \pm 0.0001 $ \\
F255W   & $1.0004 \pm 0.0001 $ \\
F300W   & $1.0000 \pm 0.0002 $ \\
F336W   & $1.0006 \pm 0.0001 $ \\
F555W   & 1.0 (by definition) \\

\end{tabular}
\end{table}

The final step is to
use the STSDAS routine {\it north\/} to determine the
orientation of the PC $y$ axis relative to celestial north, in order to
calculate the position angle of the binary.  The spacecraft roll angle, as
calculated by {\it north\/}, is
accurate to about $\pm0\degpoint04$ (S.~Casertano and E.~Nelan, 
private communications).

In the case of 14~Aur, the B component fell in the WF2 chip.  The reduction 
of its position was done with a program kindly provided by S.~Casertano, 
which takes into account the relative positions and orientations of the four 
WFPC2 chips, as well as the wavelength-dependent geometry of the four fields.
The measurement is, however, less accurate than for measurements made within 
a single chip, due to small time-dependent drifts in the relative positions 
of the CCD chips.

The largest source of uncertainty in the astrometry
arises from the centroiding of the
slightly undersampled stellar images.  Tests based
on series of dithered images of the resolved binaries Procyon and G~107-70, 
obtained 
by H.E.B. and collaborators in a separate WFPC2 program, show that the 
1$\sigma$ internal scatter in the separations as measured with {\it 
imexamine\/} at a single telescope pointing is about 
$0\secpoint003$.  The other sources of uncertainty (e.g., in the geometrical
corrections and F555W plate scale) are believed to be of order
$0\secpoint001$ or less over separations of up to several arcseconds.
The errors are, of course, larger when some of the
pixels at the center of the stellar image are saturated, as was the case for
the A component of 14~Aur and the Aa component of 56~Per. The position of
14~Aur~B relative to the A component, however, is probably an order of
magnitude less accurate, since it had to be measured across two separate
chips as noted above.

Finally, the $\zeta$~Cyg system was only marginally resolved, and we were
unable to measure the components with {\it imexamine}. Instead, we 
analyzed its image 
as follows.  First, as a reference image we chose our observation of 
HD~15638, which had been taken in the same F218W filter
under similar predicted telescope focus conditions 
(the detailed {\it HST\/} 
focus history, predicted from a model based on the instantaneous thermal status 
of the telescope, is documented at the STScI web site).  Using 
IRAF's {\it imshift\/} routine, we shifted the reference image by various 
fractional pixel amounts, and then added the shifted image to the reference 
image to simulate a partially resolved binary.  We then did a least-squares 
fit to the image ellipticities (as output by {\it imexamine}) vs.\ the known
separations of the simulated binaries. Finally, using the measured ellipticity
of the $\zeta$~Cyg image, we could read off its separation, $0\secpoint036$.
Although the formal internal error of the least-squares fit is only of order 
$0\secpoint0015$, the systematic error in the separation
introduced by the undersampling of the 
PC pixels, and by our assumption that both components of the binary are of 
equal brightness, is probably at least of order 10\%.  More sophisticated
modelling of the image might improve this result further, but is beyond 
the scope of the present paper.

The results are summarised
in Table~\ref{seps}.  Except for 14~Aur, discussed below, the position angle 
is that
of the fainter component in the UV filter used, 
relative to the brighter 
component in that filter.  It should also be noted
that the magnitude differences are
measured in the UV filters used for the observations, and are much larger at 
visual wavelengths.  Except as discussed below, 
we generally do not know which of the resolved
components is 
the WD and which is the cooler optical star. We also note that, with the 
limited photometric information available, we cannot completely 
exclude the possibility 
that the resolved object in each system is not actually a WD, but 
a normal star of similar spectral type to the target. In this case the WD
would reside in a still unresolved binary, with one of the imaged stars as its
primary. The absence of any reported radial velocity variations for these 
resolved systems (e.g. Vennes \etal \ 1998; 
except for 14 Aur C, see section 5.1.5), would tend to
rule this out but there remains a region of the parameter space (i.e.
small orbital inclination and/or orbital periods of a few years) where a WD
would remain hidden from both radial velocity and WFPC2 observations.
Observations in additional WFPC2 filters will be
necessary to definitively identify the WD components.

\begin{table}
\caption{Measurements of
the resolved binaries.}
\label{seps}
\begin{tabular}{lcccc}
Star     &  Epoch & Sep. & PA (J2000) & $\Delta$mag  \\

HD 2133 & 1999.64 & $0\secpoint602$ & $46\degpoint85$ & 1.00 \\
MS 0354.6-3650 & 2000.40 & $0\secpoint992$ & $73\degpoint44$ & 1.52 \\
HR 1358 & 1999.54 & $1\secpoint276$ & $10\degpoint62$ & 1.96 \\
56 Per Aa-Ab & 1999.79 & $0\secpoint390$ & $306\degpoint29$ & $>$3.25 \\
\phantom{56 Per} Ba-Bb & & $0\secpoint633$ & $295\degpoint88$ & 1.70 \\
\phantom{56 Per} Aa-Ba & & $4\secpoint284$ & $18\degpoint30$ & $>$4.69 \\
\phantom{56 Per} Ab-Ba & & $4\secpoint180$ & $23\degpoint40$ & 1.58 \\
14 Aur A-Cb  & 1999.73 & $15\secpoint20$: & $231\degpoint67$ & $\dots$ \\
\phantom{14 Aur} A-Ca  & & $14\secpoint26$: & $224\degpoint79$ & $\dots$ \\
\phantom{14 Aur} Ca-Cb & & $2\secpoint006$ & $290\degpoint11$ & $-0.25$ \\
\phantom{14 Aur} A-B & & $10\secpoint31$: & $8\degpoint91$ & $\dots$ \\
RE J1925$-$566  & 1999.65 & $0\secpoint217$ & $120\degpoint94$ & 0.08 \\
$\zeta$ Cyg     & 2000.39 & $0\secpoint036$: & 246: & $\dots$ \\
HD 223816  & 1999.74 & $0\secpoint574$ & $37\degpoint02$ & 0.48 \\

\end{tabular}
\end{table}

In order to facilitate such future observations, we present
images of five of our resolved binaries in Figs.~1a-1e, in which 
each frame is a $5''\times5''$ square.  The more complicated
multiple systems 56~Per and 
14~Aur are illustrated in Figs.~\ref{56perim} and ~\ref{14aurim} 
respectively.  We do not attempt to illustrate the 
marginally resolved binary $\zeta$~Cyg.

\begin{figure}
\epsfxsize=1.0\columnwidth
\leavevmode\epsffile{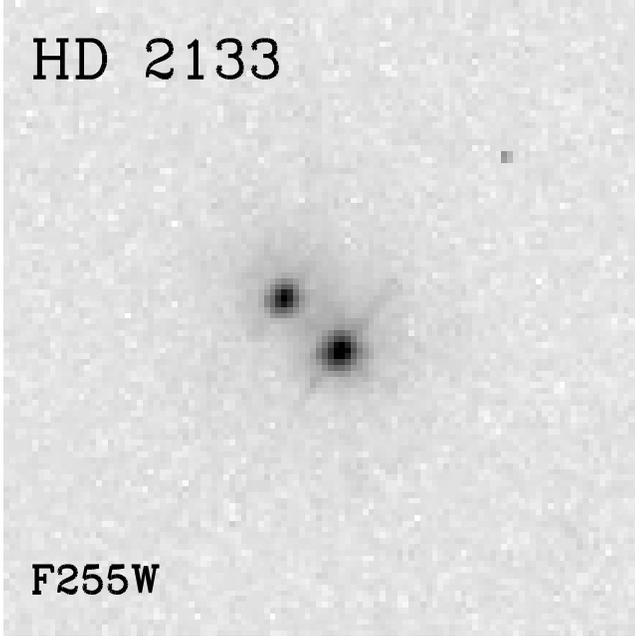}
\caption{This series of figures shows WFPC2 snapshot images of five 
binaries containing cool primary stars with resolved 
hot white dwarf companions. Each frame is $5''\times5''$ in size, and has 
north at the top and east on the left. The UV filter used is indicated in the 
lower-left corner of each image.
(a)~HD~2133; (b)~MS~0354.6$-$3650; (c)~HR~1358; (d)~RE~J1925$-566$; 
(e)~HD~223816. }
\label{images}
\end{figure}

\begin{figure}
\epsfxsize=1.0\columnwidth
\leavevmode\epsffile{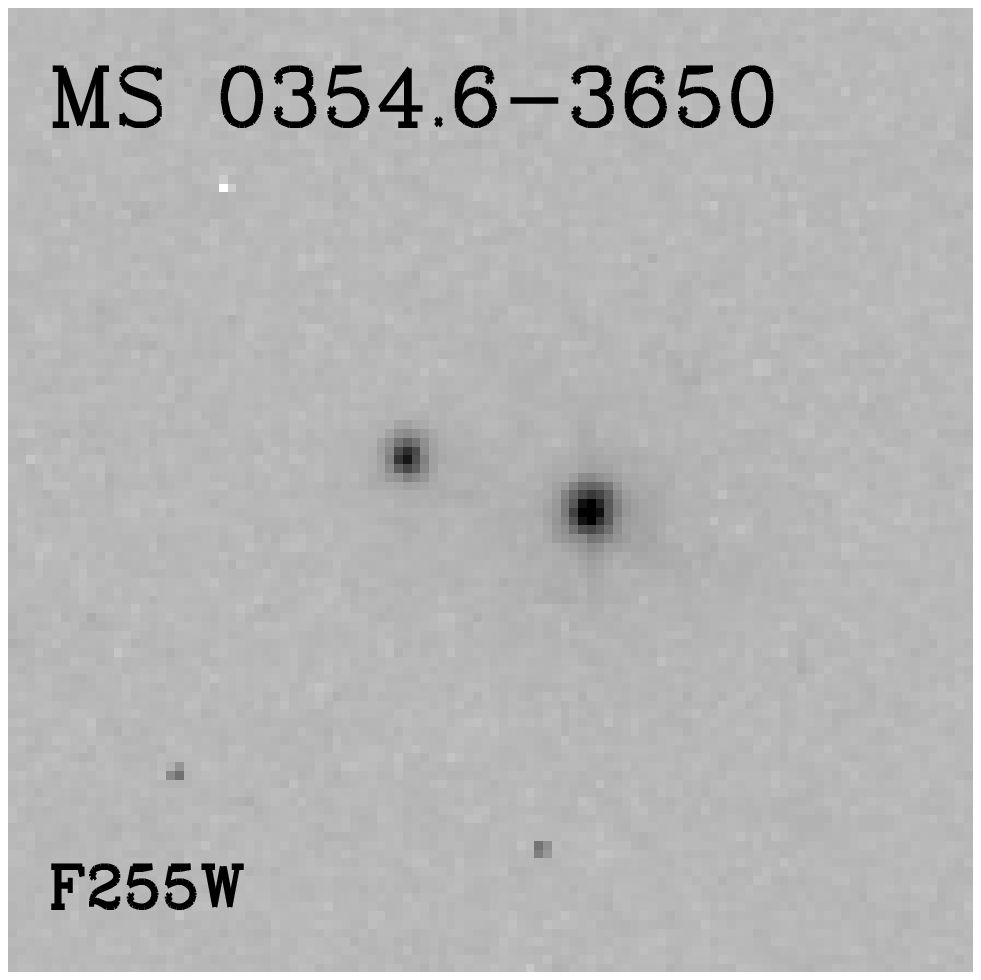}
\contcaption{(b)}
\end{figure}

\begin{figure}
\epsfxsize=1.0\columnwidth
\leavevmode\epsffile{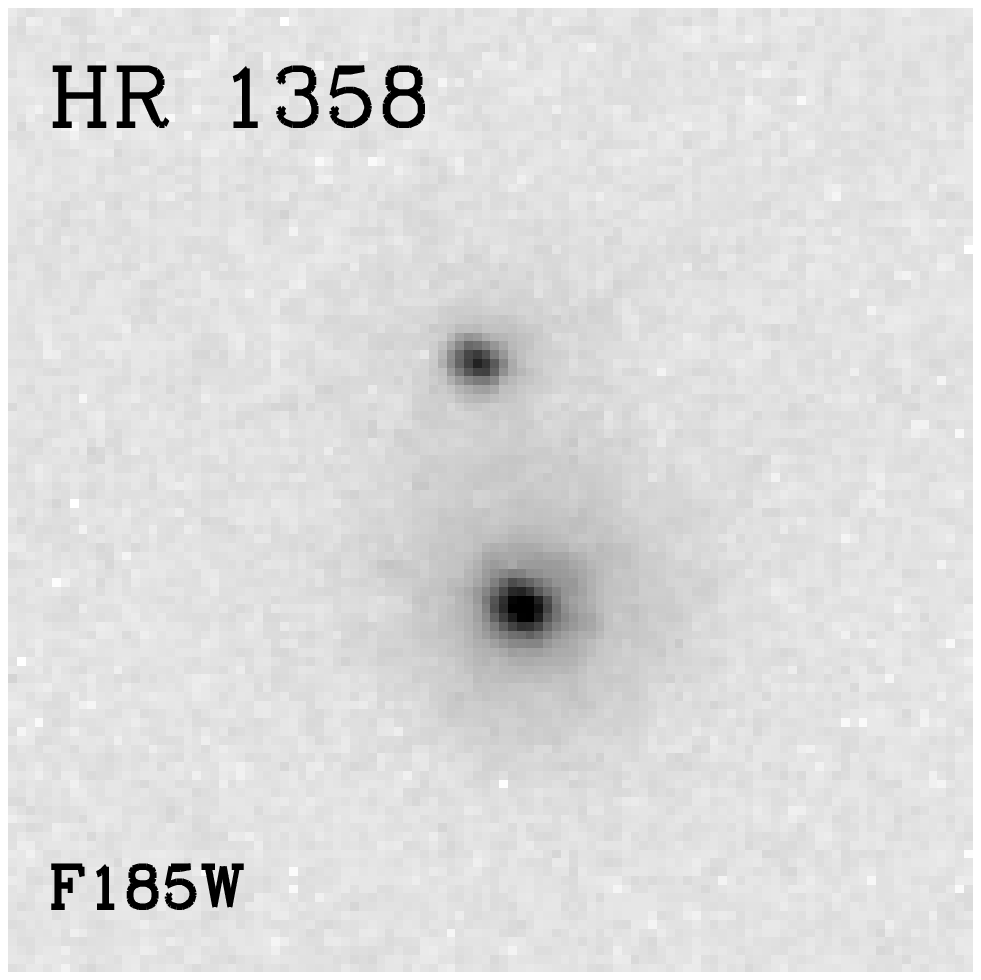}
\contcaption{(c)}
\end{figure}

\begin{figure}
\epsfxsize=1.0\columnwidth
\leavevmode\epsffile{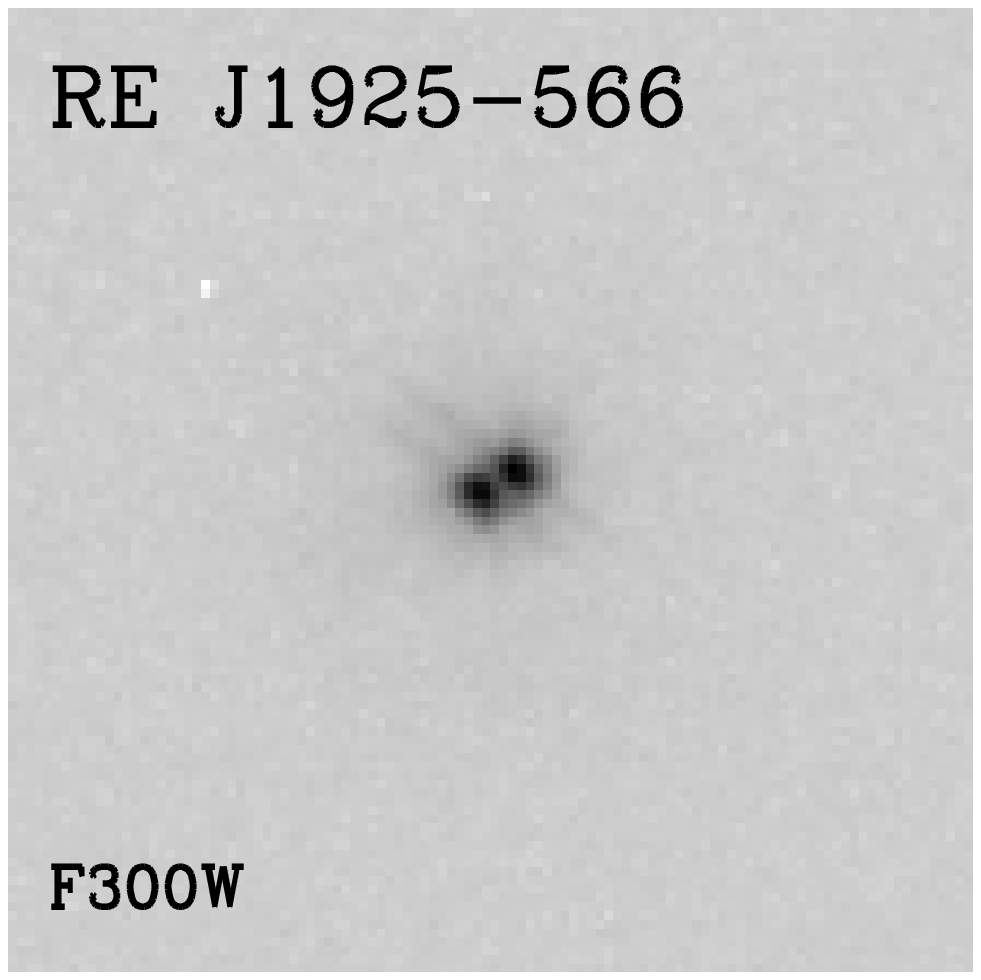}
\contcaption{(d)}
\end{figure}

\begin{figure}
\epsfxsize=1.0\columnwidth
\leavevmode\epsffile{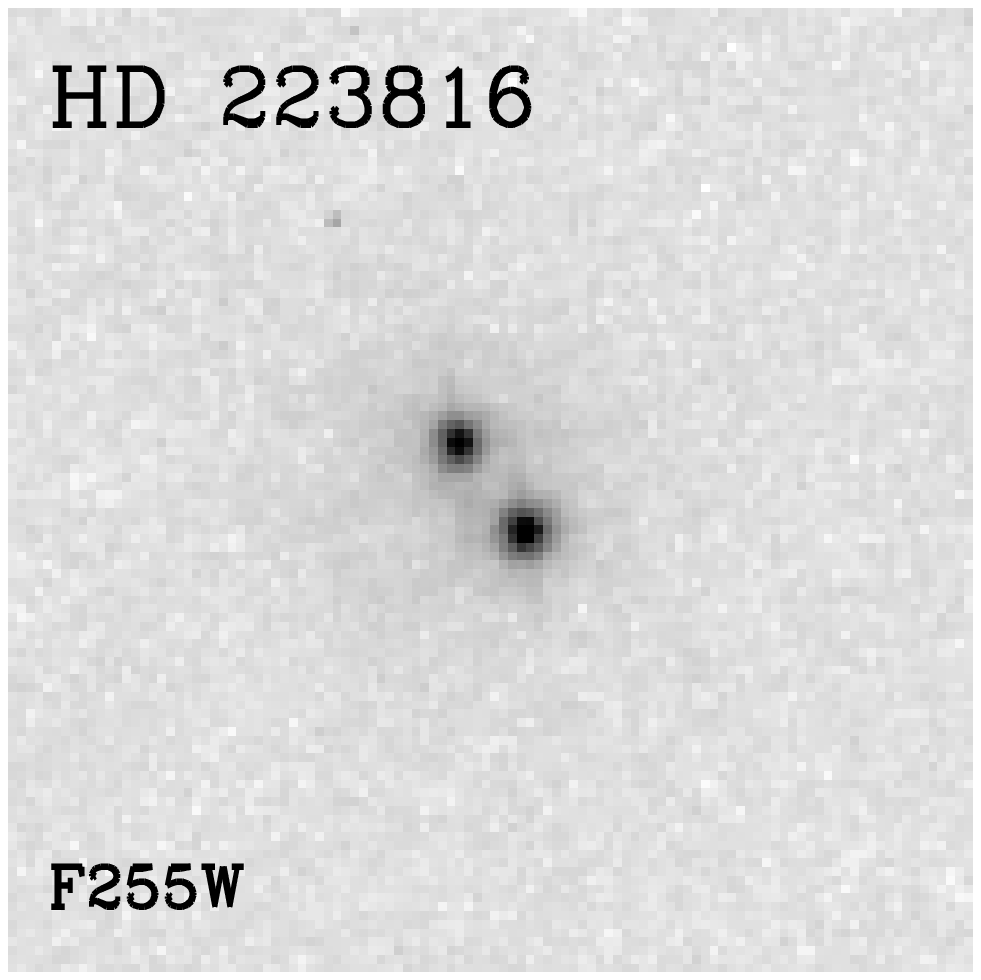}
\contcaption{(e)}
\end{figure}

From Kepler's third law we can calculate the period $P$ (in years) of
a binary system, using the usual equation
$$ P =[(a/\pi)^3/(M_{\rm pri}+M_{\rm WD})]^{1/2},$$ 
where $a$ is the semi-major axis of the orbit and $\pi$ is the parallax, both 
in seconds of arc, 
and $M_{\rm pri}$ and $M_{\rm WD}$ are the 
masses of the primary and WD,
respectively, in solar masses.

We have estimated approximate primary masses from their spectral
types, taking account of the fact that in
at least two cases (HR~1358 and 14 Aur~C) the primary is itself a binary system (see 
below).
For the WD components, we simply
assume a mass of $0.6M_\odot$
(even though some of them - see below - may be somewhat
more massive. We then calculate nominal orbital periods, 
$P_{\rm nom}$,
taking $a$ to be the measured separation listed above, and tabulate them in 
Table~\ref{pers}. In almost all
cases, the true orbital periods will of course be longer than 
$P_{\rm nom}$, since the projected
orbital separations are lower limits to the true current binary
separation and the system is most likely not presently
at maximum projected elongation; 
however, for some unusual
system geometries the true period could be less 
than $P_{\rm nom}$. The table shows
that some of the orbital periods may be
short enough for detection of the motion on timescales of a few
years or less. 

\begin{table}
\caption{Estimated primary masses and nominal orbital periods for the
resolved systems. All white dwarf masses are assumed to be $0.6 \,M_\odot$.
Stars with masses marked by ``:'' have evolved away from the main
sequence. Hence, the values tabulated are more uncertain than for the main
sequence objects.} 
\label{pers}
\begin{tabular}{lcc}
Star   & $M_{\rm pri}$ ($M_\odot$) & $P_{\rm nom}$ (yr) \\

HD 2133 & 1.1 & 590 \\
MS 0354.6$-$3650 & 1.0 & 6200 \\
HR 1358 & 2.5 & 260 \\
56 Per Aa-Ab & 1.4 & 47 \\
14 Aur Ca-Cb & 2.0 & 1307 \\
RE J1925$-$566 & 0.9 & 95 \\
$\zeta$ Cyg & 1.5: & 1.5: \\
HD 223816 & 1.1 & 290 \\

\end{tabular}
\end{table}

\subsection{Unresolved systems}

As discussed above, the remaining nine systems were unresolved.  We can 
assume that the projected separation of the components is less than about 
the FWHM of the images, which is generally $\approx$$0\secpoint08$.
Then, similarly to the above calculations for the resolved binaries, we can 
calculate nominal upper limits to the orbital periods (the true periods 
could, under special circumstances, be longer than the nominal ones, e.g. if 
the system happens to be near conjunction at the present time). These nominal 
upper-limit periods are 
tabulated in Table~\ref{unr}. The main conclusion
is that most of the systems 
could still be wide enough for the WD components to have evolved essentially 
as single stars. On the other hand, the fact that the systems are unresolved 
makes it statistically highly probable that the binaries do form a physically 
associated system, rather than a chance alignment.
Consequently --- as is likely to be the case for the resolved systems as well
--- most or all 
of them
can be used as tracers of the
initial-final mass relation, as well as in studies of the WD
mass-radius relation.

\begin{table}
\caption{Nominal upper limits on orbital periods for the unresolved systems.
All white dwarf masses are assumed to be $0.6 \, M_\odot$.
Stars with masses marked by ``:'' have evolved away from the main
sequence. Hence, the values tabulated are more uncertain than for the main
sequence objects.}
\label{unr}
\begin{tabular}{lcc}
Star   & $M_{\rm pri}$ ($M_\odot$) & $P_{\rm upper}$ (yr) \\

BD +08$^\circ$102 & 0.7 & 9.7 \\
HD 15638 & 1.2 & 47 \\
HD 18131 & 1.0: & 19: \\
63 Eri & 1.0: & 7.3: \\
HR 3643 & 2.5: & 21: \\
HD 90052 & 1.5 & 62 \\
$\beta$ Crt & 2.5: & 9.5: \\
RE J1309+081 & 1.1 & 79 \\
HR 8210 & 1.8 & 4.6 \\

\end{tabular}
\end{table}

\section{Discussion} 

We now discuss each of the observed systems briefly, starting with the 
resolved systems, and then turning to the unresolved ones.

\subsection{The resolved binaries} 

\subsubsection{HD 2133} 

Burleigh, Barstow, \& Fleming (1997)  reported the detection of HD~2133 in
the \ro\/ WFC all-sky survey and their subsequent identification of the WD 
companion
through \iu\/ observations. The $V=9.6$ F7~V primary is included in the
\hip\ catalogue, at a distance of 140~pc. Adopting this distance,
Burleigh \etal\ (1997) estimate the $T_{\rm eff}$ and $\log g$ of the WD
to be $\approx$26,800~K
and 7.75, respectively. Interestingly, the X-ray and EUV data require the
WD to have a pure H atmosphere and negligible interstellar column
density, lower than might be expected in that direction. To our knowledge, no
radial-velocity observations have been made of the system, but with a nominal
orbital period of nearly 600~yr,
it will not be
possible to detect radial velocity variations in the foreseeable future.

\subsubsection{MS 0354.6$-$3650} 

The WD was initially detected in the \ei\ Medium Sensitivity 
survey (hence the MS designation) but misidentified with a cluster of 
galaxies (Stocke \etal\ 1991). It was subsequently detected by \eu\ and 
associated with a $V=12.45$ G2~V star; an \iu\ spectrum obtained by 
Christian \etal\ (1996) confirmed the presence of a
hot WD with 
$T_{\rm eff}\ge52,000$~K. With a large physical separation and a 
nominal orbital period $\approx$6,000~years, 
we would not expect the G2~V primary to display any evidence for radial 
velocity variations. If better constraints can be placed on the WD's 
physical parameters through far-UV observations of the H Lyman series with 
\fuse , then it should be possible to show whether the two stars really 
are associated or, less likely, merely a chance alignment.

\subsubsection{HR 1358} 

The WD companion to HR~1358 (HD~27483) was originally discovered by
B\"ohm-Vitense (1993) through an \iu\/ observation of the system.  Burleigh,
Barstow \& Holberg (1998)  reanalysed the \iu\/ spectrum in association
with the \ro\/ PSPC and WFC data. The active primary of this system is a
binary in its own right, consisting of a pair of F6 V stars, with an orbital
period of 3.05~days. For the \hip\/ distance, the WD
UV spectrum is best matched by a $\log g=8.5$ and $T_{\rm eff}=22000$~K 
model, with
an estimated mass of $0.98\,M_\odot$.  As a member of the Hyades cluster, 
HR~1358 has
a known age  and it is possible to compare the age of the proposed
$6.7\,M_\odot$ progenitor with that of the cluster. Burleigh \etal\ (1998)
point out the the sum of the progenitor lifetime and WD age is
significantly less than the cluster age (0.7~Gyr), implying that the reported
WD mass is an overestimate. Now that the WD has been resolved with
\HST, it may be possible to obtain a Balmer line spectrum of it and apply the
standard analysis technique employed for isolated DAs to resolve this
problem.

\subsubsection{56 Per} 

The first indication
of a WD companion to 56 Per  
was through a flux excess in the 1565~\AA\
band of the TD-1 ultraviolet sky survey (Landsman, Simon, \& Bergeron
1996); the WD was subsequently confirmed by these authors
with an \iu\/ spectrum. 
At $T_{\rm eff}=14,000-16,000$~K,
the WD is actually too cool to be detected at EUV or soft X-ray
wavelengths. Landsman \etal\ (1996) also obtained a surface gravity of $\log
g=8.46\pm 0.2$, giving an implied WD mass of $0.9\,M_\odot$,  well
above the canonical $\approx$$0.6\,M_\odot$ typical of single DAs.

56 Per ($V=5.8$, F4~V) has a 
visual companion with $V=8.7$, known since the 19th century, at a separation 
of about $4''$. An extensive archive of measurements of the 56~Per~AB visual
binary is available at the U.S. Naval Observatory's Washington Double Star 
Catalog and was kindly provided to us
by Dr.~Brian Mason. These data show slow orbital motion, with the B component 
having moved through an arc of about $30^\circ$ since the mid-1800's.  (This 
amount of motion is roughly consistent with a nominal orbital period for the 
AB pair, 
calculated as above, of about 1500~yr.)  The B component, however, is 
unlikely to be the WD, since
Landsman \etal\ point out that 56~Per~B is 
both too bright visually to be the WD, and too distant from 56~Per~A as well, since 
their \iue\/ observation indicated that the WD is within $\sim$$1''$ of the A 
component.  

Remarkably, Fig.~2 reveals that the 
system actually has {\it 
four\/} resolved components, which can be seen in the 1700\AA \ image. 
The brightest component is seen to
have a companion at a separation of only $0\secpoint39$, and we identify this 
pair as the F4~V primary and the WD companion.  They are denoted as ``Aa'' 
and ``Ab'' in Fig.~2 (and in Table~\ref{seps}) on the assumption that Aa is 
the F star, although we caution that it is 
possible that the star labelled Aa 
could actually be the WD, and Ab the visually bright primary. 
The B component is also resolved into a close pair, which we denote Ba-Bb,
having a separation of 
$0\secpoint6$. The nominal orbital period of the Ba-Bb pair is about 100~yr.

The close F4-WD pair, with a nominal period 
near 50~yr, is rather similar to the prototype Sirius system. 
It is a prime candidate for continued \HST\/ observations, which---if a 
successor high-resolution UV imaging space 
experiment is available to continue the 
monitoring once \HST\/ has been decommissioned---would
ultimately yield
an astrometric determination of the orbit and the component masses.

\begin{figure}
\epsfxsize=1.0\columnwidth
\leavevmode\epsffile{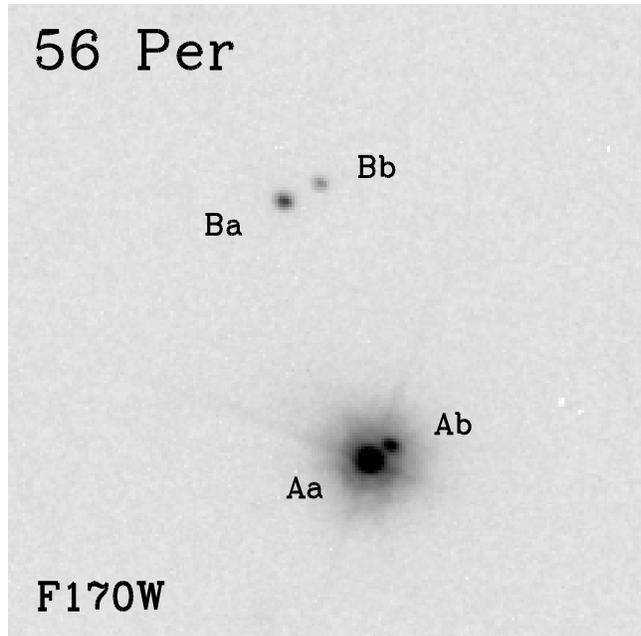}
\caption{WFPC2 image of 56 Per in the F170W filter. The image is 
$10''\times10''$ and has north at the top and east on the left. The Aa-Ab 
pair is the F4~V primary and its WD companion. The \HST\/ image reveals that 
the previously known visual companion, 56~Per~B, is also resolved into two 
components, which we label Ba and Bb.}
\label{56perim}
\end{figure}

\subsubsection{14 Aur C} 

The first of the Sirius-like binaries to be discovered from the \ro\/
WFC EUV survey (Hodgkin \etal\ 1993), 14~Aur forms an interesting
multiple system. The primary star 14 Aur~A (also cataloged as
KW~Aur) is a $\delta$~Scuti variable and a
single-lined spectroscopic binary ($P=3.79$~days; Fitch \& Wisniewski 1979)
with a spectral type of A9~IV (Morgan \&  Abt 1972) or A9~V (Abt \& Cardona 1984).  
14~Aur~B is an 11th-mag star located some 
$10''$ north of A\null. The available measurements of the AB pair from the 
Washington Double Star Catalog, along with our new \HST\/ measurement,
reveal linear relative motion, confirming that B 
is merely an optical companion.  The Washington Catalog, however,
along with our new measurement, show that 14~Aur~C shares 
a common proper motion with A, and is thus a physical 
companion.

The C component, which has been classified variously as F3~V (Halbedel 1985),
F4~V (Abt \& Cardona 1984), and F2~V (Vennes
\etal\ 1998), is also a single-lined
spectroscopic binary; its velocity variability was discovered by Webbink 
\etal\ (1992), and the orbital period was found to be 2.99~days by Tokovinin 
(1997) and, independently, by Vennes
\etal\ (1998). 
Tokovinin and Vennes \etal\ 
also found very similar $\gamma$ velocities for A and C, 
strengthening the conclusion that they form a physical (at least
quadruple) system.

Hodgkin \etal\ (1993) were able to associate the detected WD
with the vicinity of the
C component, raising the possibility that it is in fact the 2.99-day 
radial-velocity companion; if so,
14 Aur~C would be of interest as a post-common-envelope system.
However, Holberg \etal\ (1999), using \iue\/ spectra, found no evidence for
radial-velocity variations of the WD based on measurements of
metallic lines visible in the WD
photosphere; thus the WD cannot be the 2.99-day companion.

\begin{figure}
\epsfxsize=1.0\columnwidth
\leavevmode\epsffile{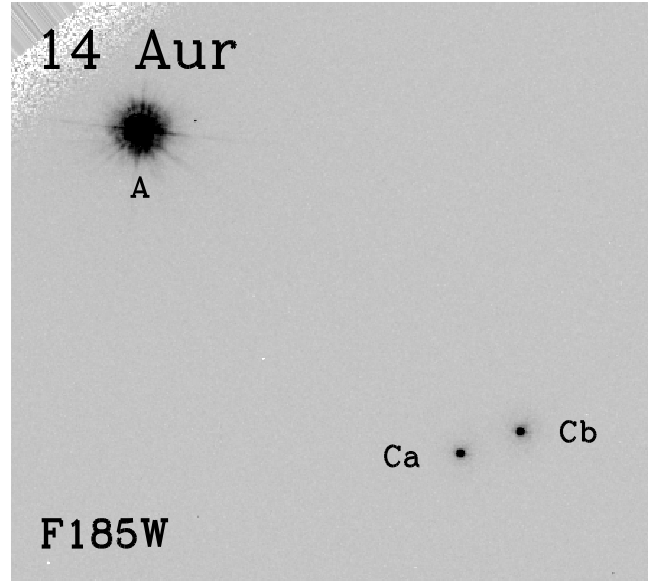}
\caption{WFPC2 image of 14 Aur in the F185W filter. The image is 
$20''$ wide and has north at the top and east on the left. The C component is 
seen to be resolved into the object previously known from the ground
(itself an unresolved binary with a 2.99d period), denoted 
Ca, and the hot WD, denoted Cb.}
\label{14aurim}
\end{figure}

The \HST\/ image, shown in Fig.~3, immediately clarifies the situation. It 
shows 14 Aur~C to be resolved into two components separated by $2''$. 
Our astrometric measurements 
(cf.\ Table~\ref{seps}), combined with the historical visual measurements, 
demonstrate
conclusively that the component labelled ``Ca'' in Fig.~\ref{14aurim}
is the F-type 
star, with a still-hidden companion, detected from radial
velocity measurements made at visual wavelengths (Webbink \etal \ 1992),.  
Hence the new
object, denoted ``Cb,'' must be 
the WD (which is actually slightly brighter in the UV frame than Ca). There 
is thus a total of
at least five components in the 14~Aur system.  If we assume that Cb is 
a physical companion of Ca (which is as yet unproven, aside from the 
statistical improbability of a chance alignment), its orbital period
about the unresolved Ca binary 
is of order 1300~yr (Table~\ref{pers}).

\subsubsection{RE J1925$-$566} 

This system, with a separation of about $0\secpoint2$,
has a nominal orbital period near 100~yr, short enough that additional \HST\/ 
imaging could 
reveal the motion within the next few years.  While Barstow
\etal\ (1994) report the detection of Ca~II H and K emission, there is no
evidence from the \ro\/ PSPC data that the G7~V primary is an X-ray source in
its own right (Vennes \etal\ 1998). The analysis of Barstow \etal\ did
suggest an unusual excess in the WFC S1 flux. However, 
the \eu\/ spectrum presented
by Vennes \etal\ indicates a higher temperature (48,000--51,000~K) for the
WD than obtained earlier from the \iu\/ Lyman~$\alpha $ line profile
($\approx$37,000~K), which explains the anomaly. Vennes \etal\ report that
they have initiated  a radial velocity study of the system. Now that we have 
resolved the WD,
we would not expect any variations to be detectable unless the G
star is itself a close binary, as might be suggested by the Ca~II emission. 

\subsubsection{$\zeta$ Cygni} 

$\zeta$ Cygni (HR~8115, HD~202109) is a prototype of the
``mild'' barium stars. The history of the
recognition of its spectroscopic peculiarities is recounted
in full detail by Griffin \&  Keenan (1992), who adopt
a classification of G8+~IIIa~Ba0.5 (abbreviated as
G8~IIIp in our table 1). Barium stars are thought to be
members of binary systems having WD companions (e.g. B\"ohm-Vitense 2000; 
Bond \&  Sion 2000; and references therein).
The abundance anomalies of the present-day primaries probably originate
in processed material transferred from the WD progenitor when it
was on the asymptotic giant branch.
The WD
companion to $\zeta$ Cygni was first identified in an \iu\ far-UV spectrum 
by B\"ohm-Vitense (1980) and confirmed by Dominy \& Lambert (1983), who
estimated its surface temperature to be $\approx$12,000~K\null. Griffin \&
Keenan (1992) showed that $\zeta$ Cygni displays radial velocity variations
with a period of $6489\pm31$~days (i.e., $\sim$18~years); we have now
marginally resolved the WD companion, and the binary separation
($\sim$$0\secpoint036$)
is not inconsistent with this orbital period. 

This system is a
prime candidate for further imaging to observe the orbital motion over
the next few years, leading to a new dynamically determined mass for the
WD. According to the spectroscopic ephemeris of Griffin \& Keenan (1992), our 
{\it HST\/} observation of $\zeta$~Cyg was taken at spectroscopic phase 0.69. 
At this phase, the radial velocity is almost exactly at the center-of-mass 
velocity, i.e., the stars should be near their smallest projected angular 
separation.  Our small measured separation agrees with this expectation.  The 
angular semimajor axis should be of order $0\secpoint19$, so at other orbital 
phases the binary should be easily resolved by {\it HST}.

\subsubsection{HD 223816} 

The WD component of this system is one of the hottest in the
sample of EUV-detected binaries, the \iu\/ data indicating $T_{\rm eff}$ to be
around 69,000~K\null.   
The EUV and soft X-ray photometry indicates that the WD
atmosphere is heavily contaminated with heavy elements; this is
confirmed by an \eu\/ spectrum (Vennes \etal\ 1998), which shows the WD 
to be
similar to G~191-B2B and RE~J0457$-$285. Vennes \etal\ detect no radial
velocity variations, consistent with the wide separation of the
system observed here. 
The presence of photospheric heavy elements
is thus likely to be due to  radiative levitation effects rather than the
presence of the wide companion. 

\subsection{The unresolved binaries} 

\subsubsection{BD $+08^\circ 102$}

The level of activity and high rotational velocity observed in the primary
star (e.g. Barstow \etal\ 1994)
suggest this is a close binary. It has been proposed
as the prototype of a new class of fast rotators, spun up by accretion
from the red-giant
wind of the WD progenitor (Jeffries and Stevens 1996). No
radial velocity variations were detected in several days of observations
(Kellett \etal\ 1995), suggesting that the period must be fairly long. Our 
nominal upper limit from the lack of resolution is about 10~yr.

\subsubsection{HD 15638} 

Both Landsman \etal\ (1993) and Vennes \etal\ (1998) classify the primary
as an F6 dwarf, but the \hip\/ distance is only compatible with
a subgiant classification. This is consistent with the F3~IV
classification in the Michigan Spectral Survey (Houk \& Cowley
1975). Vennes \etal\ report high-resolution spectra at Ca~II H and K
that show no
evidence of activity. Favata \etal\ (1995) give a rotational velocity
$v_{\rm rot} \sin i=62 \,\rm km\,s^{-1}$. 
In a short series of four measurements, Vennes \etal\
(1998) found no radial-velocity variations, with a standard deviation of
$0.7\,\rm km\,s^{-1}$, 
indicating that this is not a short-period system. Our failure to resolve the 
system with \HST\/ only constrains the orbital period to be less than about 
50~yr.

\subsubsection{HD 18131} 

The \hip\/ distance is consistent with the K0~IV-III classification of Vennes
\etal\ (1998). The primary is also an X-ray source, indicative of a short
rotation period. However, Vennes \etal's radial-velocity measurements
indicate variations of a few $\rm km\,s^{-1}$
on a one-year time scale. The implied
period is well below the nominal 19~year
limit of our observations, so it is not surprising that the system is not
resolved. Hence, it is most likely that this is also an
example of a primary spun up by
interaction with the wind of the WD progenitor. 

\subsubsection{63 Eri} 

This is a known single-lined spectroscopic binary with an orbital period of
903 days (Beavers \& Eitter 1988), about 2.5 years. We assume that the WD is 
the star in the 903-day orbit, since the period would have had to
be $\approx$7~years for the system to have been resolved by our programme. 

\subsubsection{HR 3643} 

HR~3643  is one of the two systems (the other being 56~Per) with WD 
companions
identified from
the TD-1 survey by Landsman \etal\ (1996). No evidence for radial-velocity
variations have been found for the primary, which has in fact
been adopted as a
radial-velocity standard (e.g. Layden 1994). Hence, it cannot be a short
period system. The $\approx$21~year upper
limit on the binary period imposed by
the WFPC2 observations is consistent with this. 

\subsubsection{HD 90052} 

The WD companion to HD~90052 (BD$+27^\circ 1888$) was discovered by 
Burleigh \etal\ (1997). Vennes \etal\ (1998) comment that the WD
and the F0~V primary form an unlikely pair: with a temperature between 
34,000~K--40,000~K and a surface gravity log $g=7.3-8.5$, the WD
should lie closer ($<$171~pc) than the main sequence star 
(229--275~pc, unless it is dramatically underluminous), although  
Vennes \etal\ (1998) also point 
out  that a chance alignment at this high Galactic latitude is unlikely. 
The \hs \  observation shows that the two stars are aligned within 
$\approx0\secpoint08$, and therefore very likely form a physical pair with an 
orbital period $<62$~years. A far-UV spectrum of the white dwarf, covering 
the H Lyman series, is urgently required to better constrain its 
physical parameters ($T_ {\rm eff}$, log $g$) and resolve the discrepancy  
between the distance estimates to the WD and the F0~V star. 
This can and should be carried out by \fuse.

\subsubsection{$\beta$ Crt}

Although Campbell \&  Moore (1928) reported velocity variations in this 
object, prompting Fleming \etal\ (1991) to speculate that the orbital 
period could be $<20.1$~days, both Smalley \etal\ (1997) and Vennes \etal\ 
(1998) showed that $\beta$~Crt is not a close system. In fact, 
$\beta$~Crt did show small variations from the systemic velocity during the 
15 months that it was studied by Vennes \etal\ (1998), indicating that the 
period may be of order a few years. Duemmler \etal\ (1997) also report 
that the mean radial velocity measured across 8 consecutive nights in 1997 
was significantly different from the value reported by Smalley \etal\ (1997) 
from measurements made in 1994/95, again suggesting
that the binary period 
is of order a few years. This conclusion is strengthened by 
proper motion measurements from \hip\, which indicate possible 
micro-variability in the star's motion across the sky, and suggest the 
binary period is $\sim$10~years, consistent with  
\hs 's failure to resolve the WD.

\subsubsection{RE J1309+081} 

RE J1309$+$081 is one of the few remaining {\it ROSAT\/} 
Wide Field Camera Bright
Source Catalogue (Pounds \etal\ 1993) sources without an obvious optical
counterpart. The EUV detection is almost certainly real, since it is
coincident with a similar source in the EUVE Bright Source Catalog
(EUVE~J1309$+$08.2, Bowyer et~al. 1994), but the only object in the field is
an 
inactive 11th-magnitude late F dwarf (GSC~0884.00380, F8/9~V, Burleigh et~al.
2000). Therefore, we suspected that this star might be hiding a fainter hot
WD companion, and added it to our \HST\/ 
target list accordingly, but the target was unresolved. We emphasise that 
RE~J1309+081 has yet to be confirmed as a Sirius-like binary, and the 
temperature given in Table 1 for the hypothetical white dwarf ($>$22,000K) 
merely represents the lower limit for DA white dwarfs to be detectable in 
the EUV. 

\subsubsection{HR 8210} 

HR 8210 was identified as a single-lined spectroscopic
binary system by Harper (1927), and
Vennes \etal\ (1998)
have provided an improved orbital period of $21.72173\pm 0.00005$~days by 
combining their new data
with Harper's.  Assuming that the WD is the short-period companion,
it is not surprising that we did not resolve the system.

\section{Conclusion} 

Using \HST's WFPC2 camera at UV wavelengths,
we are carrying out a systematic ``snapshot'' imaging
survey of binary systems, each comprising a
non-degenerate primary with a WD secondary. The majority of these systems
were discovered through the EUV or UV emission from the WD, which is
swamped by the overwhelming brightness of the primary star at visible
wavelengths. All of our targets
are unresolved in ground-based observations.

Of 17 systems observed to date, WFPC2 has resolved eight of them. These are 
new analogs of the well-known Sirius system.  Several of them, notably 
56~Per, $\zeta$~Cyg,
and perhaps RE~J1925+566 may have orbital periods short enough for 
the visual orbits to be determined in the foreseeable future, providing a 
significant augmentation to the small number of WDs with dynamically 
determined masses.  Two of the targets were found to be
interesting quadruple (56~Per) or 
quintuple (14~Aur) multiple systems, providing a sample of ``mini-clusters'' 
which should provide new constraints on WD evolution.  All of the resolved 
systems provide new targets for determinations of gravitational redshifts.

\section{Acknowledgements} 

The work of M.A.B. and M.R.B. was supported by PPARC, UK\null. H.E.B. and
J.B.H. acknowledge support from STScI Grant GO-8181.  J.B.H. also wishes to
acknowledge support for this work from NASA grant NAG5-3472. We are grateful
to Ray Lucas and the support staff at STScI for their assistance in
implementing this programme. We thank John Biretta and Stefano Casertano
(both STScI) for discussions of WFPC2 astrometry, and Brian Mason 
for providing data from the Washington Double
Star Catalog maintained at the U.S. Naval Observatory.
This research has made use of the SIMBAD database, operated at CDS,
Strasbourg, France.

\end{document}